\documentclass[12pt,preprint]{aastex}
\usepackage[usenames,dvipsnames]{color}
\usepackage[normalem]{ulem}

\newcommand{\nuclei}[2]{\ensuremath{\mathrm{^{#1}#2}}}

\newcommand{\proton}{\nuclei{}{p}}
\newcommand{\pt}{\proton}
\newcommand{\carbon}[1][12]{\nuclei{#1}{C}}
\newcommand{\nitrogen}[1][14]{\nuclei{#1}{N}}
\newcommand{\oxygen}[1][16]{\nuclei{#1}{O}}
\newcommand{\fluorine}[1][19]{\nuclei{#1}{F}}
\newcommand{\neon}[1][20]{\nuclei{#1}{Ne}}
\newcommand{\silicon}[1][28]{\nuclei{#1}{Si}}
\newcommand{\sulfur}[1][32]{\nuclei{#1}{S}}
\newcommand{\calcium}[1][40]{\nuclei{#1}{Ca}}
\newcommand{\iron}[1][56]{\nuclei{#1}{Fe}}
\newcommand{\cobalt}[1][59]{\nuclei{#1}{Co}}
\newcommand{\nickel}[1][58]{\nuclei{#1}{Ni}}

\newcommand{\SNeIa}{SNIa}
\newcommand{\SNIa}{SNIa}
\newcommand{\mean}[1]{\ensuremath{\langle #1 \rangle}}

\newcommand{\ye}{Y_{\rm e}}
\newcommand{\yhsi}{Y_{{\rm SiG}}}
\newcommand{\yhfe}{Y_{{\rm FeG}}}

\newcommand{\yn}{Y_{\rm n}}
\newcommand{\yp}{Y_{\rm p}}
\newcommand{\ysi}{Y_{{\rm 28Si}}}
\newcommand{\ysib}{Y_{{\rm 30Si}}}
\newcommand{\ys}{Y_{{\rm 32S}}}
\newcommand{\yar}{Y_{{\rm 36Ar}}}
\newcommand{\yca}{Y_{{\rm 40Ca}}}
\newcommand{\yfe}{Y_{{\rm 54Fe}}}
\newcommand{\ynib}{Y_{{\rm 56Ni}}}
\newcommand{\ynia}{Y_{{\rm 58Ni}}}
\newcommand{\ya}{Y_{\rm AZ}}

\newcommand\msun{\ensuremath{M_{\odot}}}	         % solar mass
\newcommand{\kB}{\ensuremath{k_{\mathrm{B}}}}	% Boltzmann constant
\newcommand{\NA}{\ensuremath{N_{\mathrm{A}}}}	% Avogadro number
\newcommand{\angstrom}{\ensuremath{\textrm{\AA}}}

\begin{document}

\title{On Silicon Group Elements Ejected by Supernovae Type Ia}

\author{
Soma De\altaffilmark{1}, 
F. X. Timmes\altaffilmark{1,2}
Edward F. Brown\altaffilmark{2,3}, 
Alan C. Calder\altaffilmark{4,5}, 
Dean M. Townsley\altaffilmark{6},
Themis Athanassiadou\altaffilmark{7}, 
David A. Chamulak\altaffilmark{8}, 
Wendy Hawley\altaffilmark{9}, 
Dennis Jack\altaffilmark{10}
}

\altaffiltext{1}{School of Earth and Space Exploration, Arizona State University, Tempe, AZ}
\altaffiltext{2}{Joint Institute for Nuclear Astrophysics}
\altaffiltext{3}{Department of Physics \& Astronomy, Michigan State University, East Lansing, MI}
\altaffiltext{4}{Department of Physics \& Astronomy, Stony Brook University, Stony Brook, NY}
\altaffiltext{5}{Institute for Advanced Computational Science, Stony Brook University, Stony Brook, NY}
\altaffiltext{6}{Department of Physics \& Astronomy, The University of Alabama, Tuscaloosa, AL}
\altaffiltext{7}{Swiss National Supercomputing Centre, Via Trevano 131, 6900 Lugano, Switzerland}
\altaffiltext{8}{Physics Division, Argonne National Laboratory, Argonne, IL}
\altaffiltext{9}{Laboratoire d'Astrophysique de Marseille, Marseille cedex 13, France}
\altaffiltext{10}{Departamento de Astronom\'\i{}a, Universidad de Guanajuato, Apartado Postal 144, 36000, Guanajuato, Mexico}

\email{somad@asu.edu}

\begin{abstract}

There is compelling evidence that the peak brightness of a Type Ia
supernova is affected by the electron fraction $\ye$ at the time of the
explosion.  The electron fraction is set by the aboriginal
composition of the white dwarf and the reactions that occur during the
pre-explosive convective burning. To date, determining the makeup of
the white dwarf progenitor has relied on indirect proxies, such as the average
metallicity of the host stellar population.  In this paper, we present
 analytical calculations supporting the idea
that the electron fraction of the progenitor systematically influences 
the nucleosynthesis of silicon group ejecta in Type Ia supernovae.  In particular, we 
suggest
the abundances generated in quasi nuclear statistical equilibrium are
preserved during the subsequent freezeout.  This allows one to
potential recovery of $\ye$ at explosion from the abundances 
recovered from an observed spectra. We show that
measurement of \silicon[28], 
\sulfur[32], \calcium[40], and
\iron[54] abundances can be used to construct $\ye$ in the silicon-rich
regions of the supernovae.  If these four abundances are determined exactly, they are sufficient
to recover $\ye$ to 6\%. This is because these isotopes dominate the
composition of silicon-rich material and iron-rich material in quasi
nuclear statistical equilibrium. Analytical analysis shows that
the \silicon[28] abundance
is insensitive to $\ye$, the \sulfur[32] abundance has a nearly linear
trend with $\ye$, and the \calcium[40] abundance has a nearly
quadratic trend with $\ye$. We verify these trends with
post-processing of 1D models and show that these trends are reflected
in the model's synthetic spectra.

\end{abstract}


\section{Introduction}\label{sec:introduction}
\label{s:introduction}

Type Ia supernovae (henceforth \SNIa) are thought to be the
evolutionary terminus for a class of binary stellar systems
\citep{whelan_1973_aa, van-den-heuvel_1992_aa, kahabka_1997_aa,
parthasarathy_2007_aa,meng_2010_aa}, the thermonuclear incineration
of one or more carbon-oxygen white dwarfs
\citep{branch_1995_aa,wang_2012_aa}, 
a primary source of iron in galaxies 
\citep{matteucci_1986_aa,tang_2010_aa, bulbul_2012_aa},
accelerators of cosmic rays and sources of kinetic energy in galaxy evolution 
\citep{wang_2011_aa,powell_2011_aa}, 
and useful tools for measuring cosmological parameters 
\citep{phillips_1993_aa, riess_1998_aa, perlmutter_1999_aa, kowalski_2008_aa,
wood-vasey_2008_aa, hicken_2009_aa, riess_2011_aa,
conley_2011_aa,foley_2011_aa,foley_2011_ab, sullivan_2011_aa,
suzuki_2012_aa, silverman_2012_ac, silverman_2012_ab, silverman_2012_aa}.

The peak luminosity of \SNeIa\ is set by
the radioactive decay chain $\nickel[56]\to\cobalt[56]\to\iron[56]$ \citep{arnett_1979_aa,Colgate1980The-Luminosity-,arnett_1982_aa,arnett_1985_ab}, and the observed photometric
correlation between the peak luminosity and the timescale over which
the light curve decays from its maximum \citep{phillips_1993_aa} is
understood physically as having both the luminosity and opacity being
set by the mass of \nickel[56] synthesized in the explosion
\citep{arnett_1982_aa,pinto_2000_aa, mazzali_2006_aa, kasen_2007_aa}.
When corrected for the correlation between peak luminosity and light
curve decay timescale, the
 intrinsic dispersion in \SNIa\ distances is $\sim$0.14 mag 
\citep[ $\le$ 7\% in distance,][]{jha_2007_aa}, not all of
which can be attributed to statistical error. This correction removes
the dispersion that is attributed primarily from the difference in
\nickel[56] mass and other physical effects.  For example, the
residual dispersion in the Hubble diagram is reduced by excluding
those \SNIa\ with high-velocity ejecta
\citep{foley_2011_aa,foley_2011_ab,foley_2012_aa}.  Furthermore, there
appears to be a need to correct for the host galaxy
\citep{howell_2011_aa} as some properties of the host stellar
population are apparently imprinted on the explosion.  Accounting for
such systematic effects potentially allows for more accurate
determinations of the distance modulus from the observed light curve
and spectra.  This implies identifying and extracting observable
physical effects that may create a dispersion between the \SNIa\ light curves.

Over the last decade a number of observational and theoretical
studies have sought to uncover such systematic effects 
from variables other than \nickel[56], with the aim of making distance measurements 
more precise and improving our quantitative understanding of 
the progenitor systems.
Examples include 
the empirical correlations between the spectra and light curve
\citep{barbon_1990_aa, branch_1993_aa, nugent_1995_aa,
blondin_2006_aa, bongard_2006_aa, branch_2009_aa,
nordin_2011_aa, blondin_2011_aa, foley_2011_aa,foley_2012_aa},
the dependence of the peak brightness on the progenitor metallicity
\citep{hoeflich_1998_aa, umeda_1999_aa, timmes_2003_aa, travaglio_2005_aa,
roepke_2006_aa, ellis_2008_aa, gallagher_2008_aa,
piro_2008_aa, chamulak_2008_aa, badenes_2008_ab, howell_2009_aa,
neill_2009_aa, townsley_2009_aa, sullivan_2010_aa, jackson_2010_aa,
bravo_2011_aa,foley_2013_aa},
asymmetries in the explosion 
\citep{howell_2001_aa, kasen_2003_aa, kasen_2004_aa, wang_2008_aa,
kasen_2009_aa, chamulak_2012_aa}, 
central density and carbon-oxygen ratio 
\citep{hoeflich_1998_aa, dominguez_2001_aa,
roepke_2006_aa}, 
age of the progenitor
\citep{scannapieco_2005_aa,mannucci_2006_aa,sullivan_2006_aa, sullivan_2010_aa, krueger_2010_aa, krueger_2012_aa}, 
abundance ratios of neutron-rich isotopes to \nickel[56] 
\citep{mazzali_2006_aa}, 
and the opacity of the overlying material
\citep{mazzali_2001_aa, kasen_2007_aa}.
A consensus is still lacking, however, on the progenitor systems 
as well as on how differences in initial conditions create variances 
in the observed properties of \SNIa.

The composition of the white dwarf should have an effect on the
nucleosynthesis during the explosion and thus on the
isotopic abundances of the final composition.  For example, most of a main-sequence
star's initial metallicity comes from the CNO and \iron[56] nuclei
inherited from its ambient interstellar medium.  The slowest step in
the hydrogen burning CNO cycle is proton capture onto
\nitrogen[14]. Consequently, all catalyst nuclei are converted to
\nitrogen[14] when hydrogen core burning on the main sequence is
completed. During helium core burning the reaction sequence
$\nitrogen(\alpha,\gamma)\fluorine[18](\beta^{+}\nu_{e})\oxygen[18](\alpha,\gamma)\neon[22]$
converts most of the \nitrogen[14] into \neon[22].  From this point
forward, stars have a net positive neutron excess $\eta$, defined as
$\eta=1-2\mean{Z}/\mean{A}=1-2\ye$ where $\mean{Z}$ is the mean
atomic number, $\mean{A}$ is the mean nucleon number, and $\ye$ is
the electron fraction.  Additional burning stages will be driven
towards producing more neutron-rich elements at the expense of other
elements.

For example, this net neutron excess increases the production
of neutron-rich isotopes such as \iron[54] and \nickel[58] instead of
radioactive \nickel[56] in the regions of the white dwarf that produce
most of the iron group isotopes during the explosion.  This leads to a
linear correlation between the birth metallicity and the peak
brightness \citep{timmes_2003_aa,roepke_2006_aa,foley_2013_aa}.
However, the range of birth metallicities $Z_{{\rm birth}}$ is not
large enough to account for the full diversity of \SNIa\ peak
luminosity
\citep{gallagher_2005_aa,gallagher_2008_aa,howell_2009_aa}. This has
encouraged exploration of other factors that may impact neutronization
prior to the explosion. Some explorations have focused on the
$\sim$1000 yr long convective simmering of the white dwarf
prior to explosion.  The
convective region is driven by the $\carbon[12]+\carbon[12]$ reaction
and extends outward in mass from from the core to $\approx 1.2 \msun$ \citep{piro_2008_ab}.
During this simmering phase the reaction sequence
$\carbon[12](\pt,\gamma)\nitrogen[13](e^-,\nu_e)\carbon[13]$ increases
 the neutron excess by an amount
that depends on the total mass of carbon burned prior to the explosion
\citep{piro_2008_aa,chamulak_2008_aa}.  Such studies have established
the existence of a ``floor'' level of neutronization that is larger
than the neutronization due to the birth metallicity when 
$Z_{{\rm  birth}}\lesssim 2/3 \ Z_{\sun}$, where $Z$ is the
metallicity.  Thus, simmering may mask correlations between
\SNIa\ properties and the birth metallicity. For the purpose
of this paper, however, what matters is that the white dwarf has a well defined
$\ye$ when it explodes, and not how the white dwarf achieved that $\ye$
distribution.

On the other hand, observational \SNIa\  surveys exploring the impact of potential
metallicity effects invariably use the metallicity of the host galaxy
as a proxy for the metallicity of the progenitor white dwarf
\citep{ellis_2008_aa,gallagher_2008_aa,howell_2009_aa, neill_2009_aa,
  sullivan_2010_aa}.  It is well established, however, that there is a
relatively large scatter in stellar iron to hydrogen ratios, $\Delta[\mathrm{Fe/H}]
\sim 0.5\,\mathrm{dex}$, at any given age for stars in the Milky Way
\citep{twarog_1980_aa, edvardsson_1993_aa, chen_2000_aa,
  feltzing_2001_aa, rolleston_2000_aa, pedicelli_2009_aa}.  For
example, \cite{feltzing_2001_aa} constructed an age-metallicity
diagram for 5828 dwarf and sub-dwarf stars from the Hipparcos Catalog
using evolutionary tracks to derive ages and Str\"omgren photometry to
derive metallicities.  They conclude that the age-metallicity diagram
is well-populated at all ages, that old but metal-rich stars exist,
and that the scatter in metallicity at any given age is larger than
the observational uncertainties.  Alternatively, by following the
chemical evolution of homogeneous galaxy models with the evolution of
the supernova rates in order to evaluate the metallicity distribution
function, \citet{bravo_2011_aa} find the mean metallicity of
\SNIa\ and the metallicity of the host galaxy are tightly correlated
when both metallicities are measured as the CNO abundance.

If the composition of the white dwarf has an observable effect on the
\nickel[56] production and thus the \SNIa\ light curve, it could have
an effect on other elements as well.  In this paper we present a new direct
method to measure the electron fraction $\ye$ in the silicon-rich
regions for individual \SNeIa\  by using observed abundances of 
Si, S, Ca, and Fe. Our method follows from
the twin facts that \silicon[28], \sulfur[32], \calcium[40] and
\iron[54] are produced in a quasi nuclear statistical equilibrium
(henceforth QNSE) environment, and that the abundance levels achieved
during QNSE do not change during the subsequent freezeout as the
\SNIa\ expands. Thus, the QNSE abundance levels of these elements are
recorded in the spectra. Working in reverse, from the
observed abundances we can apply the QNSE relations to determine the
QNSE abundances and thus determine a reasonably accurate measure of
$\ye$ in the silicon group producing regions.  This method is
independent of any \SNIa\ explosion model and assumes only that the
isotopes are synthesized in a QNSE state. Our primary motivation behind
constraining $\ye$ is to reduce the residual
dispersion in the Hubble diagram by correcting for a potentially
measurable systematic effect.  Our secondary aim is to provide
rigorous nucleosynthesis constraints that can guide the modeling of
\SNIa\ synthetic spectra.

Our paper is organized as follows. In \S\ref{sec:constructing} we
establish the QNSE equations that connect the abundances to $\ye$ and
present a method for constructing the electron fraction from the
\silicon[28], \sulfur[32], \calcium[40] and \iron[54] abundances.  In
\S\ref{sec:essential} we show the trends predicted by our QNSE based
relations are present in the nucleosynthesis and spectra of common 1D
\SNIa\ models.  Finally, in Section \ref{sec:implications} we discuss
the implications of our results.

\section{Constructing $\ye$ from measured abundances}
\label{sec:constructing}

In this section we develop a framework based on equilibrium
thermodynamics and the conservation laws that allows construction of
the electron fraction $\ye$ from the major abundances in QNSE
silicon-rich material.  We then conclude this section by showing the
principle functional dependencies of the silicon group (henceforth
SiG) and iron group (henceforth FeG) on $\ye$.

\subsection{Basic framework}
\label{s:theory}

We first select a system consisting of the major SiG and FeG elements
to trace out the most useful equations connecting the individual
abundances and their relationship to $\ye$.  We choose \silicon[28],
\sulfur[32], and \calcium[40] from the SiG isotopes, and \nickel[58]
and \iron[54] from the FeG isotopes. As Figure \ref{fig:freeze}
 suggests, 
these are the dominant isotopes under QNSE conditions.
Conservation of mass and charge can therefore be expressed as
\begin{eqnarray}
\yn + \yp  + 28\ysi + 32\ys + 40\yca + 54\yfe + 58\ynia &=&1 \label{e1final} \\
\yp + 14\ysi + 16\ys + 20\yca + 26\yfe + 28\ynia &=& \ye 
\label{e2final}
\end{eqnarray}
From minimization of the Helmholtz free energy there follows the
fundamental QNSE relations
\citep{bodansky_1968_aa,hix_1996_aa,meyer_1998_aa,iliadis_2007_aa}
\begin{eqnarray}\label{qse}
\frac{Y_{{A,Z}}}{Y_{{A',Z'}}} &=& f(\rho,T)\yp^{{ Z-Z'}} \yn^{{ A-A'-(Z-Z')}}, \\
f(\rho,T) &=& \frac{G_{{ A,Z}}}{G_{{ A',Z'}}}\left (\frac{\rho
\NA }{\theta}\right )^{A-A'}\exp\left(\frac{B-B'}{\kB T}\right), \label{e.fdef} \\
\theta &=& \left(\frac{m_{\mathrm{u}}\kB T}{2\pi \hbar^{2}}\right)^{\frac{3}{2}} .
\label{e.theta}
\end{eqnarray}
Here $T$ is the temperature, $\rho$ is the baryonic mass density,
$G_{A,Z}$ is the temperature-dependent partition function, $B$ is the
nuclear binding energy, $\NA$ is the Avogadro constant, $\kB$ is the
Boltzmann constant, and $m_{\mathrm{u}}$ is the atomic mass unit.  The molar abundances are the local
abundances that correspond to a region of the star associated with a
specific $\rho$ and $T$.
The nuclei are
treated as an ideal gas and we ignore screening corrections, both of
which are justifiable assumptions under the thermodynamic conditions
of interest.
Specification of $T$, $\rho$, $\ye$, the aggregate molar abundance of
the SiG isotopes $\yhsi$, and the aggregate molar abundance of the FeG
isotopes $\yhfe$ is sufficient to solve for all the abundances in a
two-cluster QNSE environment.
At a given $\rho$ and $T$, we use Equation~(\ref{qse}) to write $Y_{\rm 32S}$ and
$Y_{\rm 40Ca}$ in terms of $Y_{\rm 28Si}$, $Y_{\rm p}$, and $Y_{\rm n}$.
Similarly $Y_{\rm 58Ni}$ is written in terms of $Y_{\rm 54Fe}$, $Y_{\rm p}$,
and $Y_{\rm n}$.  This leaves us with four unknowns, $Y_{\rm p}$, $Y_{\rm
n}$, $Y_{\rm 28Si}$, and $Y_{\rm 54Fe}$, and, for a known $\ye$, four
constraints: Equations~(\ref{e1final}) and (\ref{e2final}), and the sums $Y_{\rm SiG} = Y_{\rm
28Si}+Y_{\rm 32S}+Y_{\rm 40Ca}$ and $Y_{\rm FeG}=Y_{\rm 54Fe}+Y_{\rm 58Ni}$,
which are both specified externally to the solution of the QNSE.

Measurement of four quantities $\ysi$, $\ys/\ysi$, $\yca/\ys$,
and $\yfe/\ysi$ is an equally sufficient basis from which to solve for
all the abundances in the silicon-rich region of \SNIa.
For our choice of isotopes, Equation~(\ref{qse}) leads to 
\begin{eqnarray}
\label{eqqnse1}
\frac{\ysi}{\ys} & \approx &  \left(\frac{\rho\NA}{\theta}\right)^{-4} \yp^{-2} \yn^{-2} \exp\left(\frac{B_{\mathrm{28Si}}-B_{\mathrm{32S}}}{\kB T}\right) \\
\label{eqqnse3}
\frac{\yfe}{\ynia} & \approx &  \left(\frac{\rho\NA}{\theta}\right)^{-4} \yp^{-2} \yn^{-2} 
\exp\left(\frac{B_{\mathrm{54Fe}}-B_{\mathrm{58Ni}}}{\kB T}\right)\\
\label{eqqnse2}
\frac{\ys}{\yca} & \approx &  \left(\frac{\rho\NA}{\theta}\right)^{-8} \yp^{-4} \yn^{-4}
\exp\left(\frac{B_{\mathrm{32S}}-B_{\mathrm{40Ca}}}{\kB T}\right). 
\end{eqnarray}
Here we assume all ratios of nuclear partition functions are
unity.  This is justifiable, as at typical QNSE temperatures the nuclei are mostly in their ground state, and all of these nuclei have zero spin. 

Using Equations~(\ref{eqqnse1}) and (\ref{eqqnse2}), consider the local
SiG element ratio
\begin{eqnarray}
\Phi (T) &=&\frac{\ysi}{\ys} \left( \frac{\yca}{\ys} \right )^{1/2}\nonumber\\ 
 &\approx& \exp \left( {\frac{B_{{\rm 28Si}}-B_{{\rm 32S}}-0.5(B_{{\rm 32S}}-B_{{\rm 40Ca}})}{\kB T}} \right)\nonumber\\
  &=& \exp\left(\frac{-1.25}{T_{9}}\right),
\label{e:phi}
\end{eqnarray}
where $T_{9}$ is the temperature in units of $10^{9}\,\mathrm{K}$.
Typical temperatures in the QNSE regions where the SiG elements 
are formed are $(3.4\textrm{--}4.0)\times10^{9}\,\mathrm{K}$ (a range of 15\% in
temperature); over this range $\Phi$ varies by 6\%, from 0.73 to 0.69. 
Measuring $\Phi$ at a single epoch from the abundance ratios
$\ysi/\ys$ and $\yca/\ys$ allows a test of whether the SiG material
was produced in a QNSE state.  With sufficient precision, measurement
of $\Phi$ allows a determination of the temperature in the QNSE region
before freeze-out.  We assume for the remainder of this paper that
such precision is available and that the QNSE temperature is a known
quantity. Measuring $\Phi$ at multiple epochs when silicon features
dominate the \SNIa\ spectrum allows trends in the QNSE temperature to
be assessed.

More generally, a double ratio of the form 
\begin{equation}
K  = \frac{Y_{Z-2,A-4}}{Y_{Z,A}}\frac{Y_{Z'+2,A'+2}}{Y_{Z',A'}}\approx \exp\left[\frac{(B_{Z-2,A-2}-B_{Z,A}) - 
 (B_{Z',A'} -B_{Z'+2,A'+4} )}{\kB T}\right]
\end{equation}
is independent of $\rho$, $\theta$, $\yp$, and $\yn$. 
 Additionally, if the isotopes in this ratio are major constituents of
clusters in NQSE, then the argument of the exponential will be of
order unity and $K$ will not vary strongly over the narrow range of
temperature for which NQSE conditions attain. A relatively precise
value of $K$ can then be specified from the ratio $\Phi$.  Such a
quasi-constant can  also be defined for the FeG
elements. Equations~(\ref{eqqnse1}) and (\ref{eqqnse3}) imply that
\begin{equation}
\Psi \equiv \frac{\ynia}{\yfe}\frac{\ysi}{\ys}
\approx \exp\left(\frac{6.36}{T_{9}}\right),
\label{e:psi}
\end{equation}
where 
we again assume all ratios of nuclear partition functions are unity, and, by construction, $\Psi$ is independent of $\rho$ and $\theta$.  Over the range of QNSE temperatures, $\Psi$ 
varies by 28\%: from $\Psi=6.5$ at $T_{9}=3.4$ to $\Psi = 4.9$ at
$T_{9}=4.0$. 

\subsection{A recipe to construct $\ye$ from the major elements}
\label{recipe}

For our simplified system consisting of the a few major SiG and FeG
elements, Equation~(\ref{e2final}) may be written as
\begin{equation}
\ye  = \ysi \left [ 14 + 16 \frac{\ys}{\ysi} + 20 \frac{\yca}{\ys}\frac{\ys}{\ysi} + 26\frac{\yfe}{\ysi} 
+ 28 \frac{\ynia}{\ysi} + \frac{\yp}{\ysi} \right ] .
\label{eq:ye_major}
\end{equation}
We factor out $\ysi$ because, as we show in
\S\ref{sec:essential}, the silicon yield is the least sensitive to
changes in $\rho$, $T$, and $\ye$ in QNSE material. We may also 
drop $\yp$ since it much smaller ($\yp < 10^{-4}$) than
the other abundances.

The first step in reconstructing $\ye$ is to determine 
from observations the ${\ysi}/{\ys}$ and $\yca/\ys$ abundance ratios
from strata with similar velocities.  Measurement of these ratios determines the second and third terms of Equation~(\ref{eq:ye_major}).  Their ratio also forms $\Phi$ (Equation~(\ref{e:phi})), which if near unity  verifies that the SiG elements were
synthesized in a QNSE environment.  This relation may also be inverted
to determine the temperature of the QNSE environment when the SiG
elements were synthesized.  

The next step is to measure the $\yfe$/$\ysi$ abundance ratio.
Usually it is difficult to extract the \iron[54] abundance from the
iron lines. However, \iron[54] is the only iron isotope that is
abundant in the regime where both \silicon[28] and \sulfur[32] are
also abundant, in the absence of significant mixing of the QNSE
material with core material.  The reason for this is that in NSE, where most of the mass is in the
iron group, the requirement that $Z \approx A$ forces \nickel[56] to be
the dominant abundance. In contrast, for QNSE, most of the mass is in the Si-group
isotopes and this charge/mass constraint is lifted, so that the
greater binding energy of the slightly neutron-rich \iron[54] results
in \iron[54] having a larger abundance than \nickel[56]
\citep{hix_1996_aa,meyer_1998_aa,iliadis_2007_aa}.  In the absence of
large-scale mixing the \iron[54] produced in the silicon-rich regions
is physically separated from the \nickel[56] produced deeper in the
core, so that signatures of iron  at early times from \iron[54] do not depend
on the \nickel[56] decay chain. Therefore, if iron features are detected in the early time spectra
($\approx 8\,\mathrm{d}$) at the same expansion velocities where SiG elements
dominate the spectral features, they are produced by \iron[54]. This result is due to
material being in QNSE and is not dependent on any particular
\SNIa\ model.

The final step is to determine the $\ynia/\ysi$ abundance ratio.
Using Equation~(\ref{e:psi}) we write
\begin{equation}
\label{e:fe54}
\frac{\ynia}{\ysi} = \frac{\ynia}{\yfe}\frac{\yfe}{\ysi} = 
\Psi \frac{\ys}{\ysi}\frac{\yfe}{\ysi} .
\end{equation} 
Thus, the last two terms of Equation~(\ref{eq:ye_major}) are
determined and may be rewritten as
\begin{equation}
\ye  = \ysi \left [ 14 + 16 \frac{\ys}{\ysi} + 20 \frac{\yca}{\ys}\frac{\ys}{\ysi} + 26\frac{\yfe}{\ysi} 
+ 28 \Psi  \frac{\ys}{\ysi}  \frac{\yfe}{\ysi} \right ] .
\label{eq:ye_major03}
\end{equation}
Equation (\ref{eq:ye_major03}) is our principal result.  
 Highly
accurate measurement  abundance determinations
of the four quantities $\ysi$, $\ys/\ysi$, $\yca/\ys$, and $\yfe/\ysi$ under
the relevant temperatures is sufficient to determine $\ye$ to within 6\% because
these abundance dominate the QNSE composition.

\subsection{Including non-major elements to refine the $\ye$ estimate}
\label{sec:non-major}

The abundance of any $\alpha$-chain SiG element, which can be used to improve the accuracy
of the $\ye$ determination, can be recovered using
Equation~(\ref{qse}) and ${\ys}/{\ysi}$. For example, the abundance ratio ${\ys}/{\yar}$ is related to
${\ysi}/{\ys}$ by
\begin{equation}
\frac{\ysi}{\ys} = K_2 \frac{\ys}{\yar} 
\label{e:argon}
\end{equation}
with $K_2\approx\exp(-3.56/T_{9})$.  As the temperature ranges from
$T_{9}=3.4$ to $4.0$, $K_{2}$ varies from
0.35 to 0.41, a variation of 16\%.  Including $\yar$ in the sum
for $\ye$ in Eq.~(\ref{eq:ye_major}), adds the term $36 K_2(\ys/\ysi)^2$
to the right hand side of Eq.~(\ref{eq:ye_major03}). Other 
$\alpha$-chain non-major elements may be added in a similar manner.

Deviation of $\ye$ from 0.5 is primarily due to the major element $\yfe$ 
with contributions from other non-major SiG and FeG elements. For a more accurate
determination of $\ye$, one can use the mass conservation, Equation~(\ref{e1final}), 
to determine the abundances of these non-major
isotopes. For example, consider the case when $\ynib$ and
$\ysib$ are to be included. Treating $\yp$ and $\yn$ as 
trace abundance in the QNSE regions, Equation~(\ref{e1final}) becomes
\begin{equation}
\ysi \left [ 28 + 32\frac{\ys}{\ysi} + 40\frac{\yca}{\ysi} + 54\frac{\yfe}{\ysi} 
+ 58\frac{\ynia}{\ysi} \right ] + 
\ynib \left [ 56 + 30\frac{\ysib}{\ynib} \right ] = 1 
\enskip ,
\label{e:minor02}
\end{equation}
which we are going to solve for $\ynib$ in the QNSE region.
From Equation~(\ref{qse}) the abundance ratio $\ysib/\ynib$ may
be written as 
\begin{equation}
\frac{\ysib}{\ynib}=K_{3}\frac{\ysi}{\ynib}\frac{\ynia}{\ynib}
= K_{3} \left (\frac{\ysi}{\ynib} \right )^2 \frac{\ynia}{\ysi} 
\enskip ,
\label{e:minor03}
\end{equation}
where $K_{3} \approx \exp(-39.3/T_{9})$.
Substituting Equation~(\ref{e:minor03})
into Equation~(\ref{e:minor02}) and using previous relations gives
\begin{eqnarray}
\lefteqn{56 \ynib + 30 K_{3} \frac{\ysi^{2}}{\ynib}\left( \Psi \frac{\ys}{\ysi} \frac{\yfe}{\ysi} \right)  = 
1-\ysi \times} && \nonumber\\
&& \left [ 28 + 32\frac{\ys}{\ysi} + 40 \frac{\yca}{\ys}\frac{\ys}{\ysi} 
+ 54\frac{\yfe}{\ysi}
+ 58\Psi \frac{\ys}{\ysi} \frac{\yfe}{\ysi} \right ] .
\label{e:minor04}
\end{eqnarray}
Multiplying Equation~(\ref{e:minor04}) by $\ynib$ 
thus yields a simple quadratic equation, 
$56 \ynib^2 - b \ynib + c = 0$,
with $c = 30 K_{3}\ysi^2 \ \ynia/\ysi$ and $b$ being the right hand side of Equation~(\ref{e:minor04}).
Taking $\ynib$ as the positive root and substituting it into
Equation~(\ref{e:minor03}) then gives $\ysib$.
The derived $\ynib$ and $\ysib$ abundances may then be used 
to refine the estimate for $\ye$ by adding the now known terms to 
Equation~(\ref{eq:ye_major03}):
\begin{eqnarray}
\ye  &=& \ysi \left [ 14 + 16 \frac{\ys}{\ysi} + 20 \frac{\yca}{\ys}\frac{\ys}{\ysi} + 26\frac{\yfe}{\ysi} 
+ 28 \Psi \frac{\ys}{\ysi}  \frac{\yfe}{\ysi} \right ] \nonumber \\
&& +  \ynib \left [ 28 + 14\frac{\ysib}{\ynib} \right ] .
\label{eq:ye_minor06}
\end{eqnarray}
Abundances of other non-major elements
may be added in a similar manner to the example given above
to improve the determination of $\ye$.

\subsection{On the expected abundance trends with $\ye$}
\label{sec:trends}

In this section we section we seek a simple analytic relation between 
the abundances of the major QNSE elements, $\ysi$, $\ys$ and $\yca$ with
respect to $\ye$. We begin by re-writing Equation~(\ref{qse}) as
\begin{eqnarray}
\frac{1}{f(\rho,T)}\frac{Y_{AZ}}{Y_{A'Z'}}= \yn^{A-A'-(Z-Z')}\yp^{Z-Z'} .
\label{eqmath0}
\end{eqnarray}
We may assume without loss of generality that $A>A'$ and $Z>Z'$. Now let
\begin{eqnarray}
w = \sum_{i \ne {\rm protons}} Z_{i} Y_{A_{i}Z_{i}} , 
\label{eq:w}
\end{eqnarray}
which is identical to the definition of the electron fraction $\ye$ but
without the free protons.  We define, $v=Y_{A'Z'} f(\rho,T)$. Therefore,
with this notation Eqn.~\ref{eqmath0} becomes, 
\begin{equation}
Y_{A_iZ_i} = v(1-\ye - w)^{A_i-A'-(Z_i-Z')}(\ye - w)^{Z_i-Z'}  .
\label{eqmath2}
\end{equation}
Multiplying by $Z_{i}$ and summing,
\begin{equation}
w=\sum_{i \ne {\rm protons}} Z_i Y_{A_i,Z_i} =\sum_{i \ne {\rm protons}} Z_{i}v_{i}
(1-\ye - w)^{(A_i-A')-(Z_i-Z')}(\ye - w)^{Z_i-Z'}  .
\label{eqmath3}
\end{equation}

Since $\ye$ $\rightarrow$ 0.5, and $\ye < 0.5$, therefore,
$0<(1-\ye-w)<1$ and also that $0<(\ye-w)<1$. This imples that the RHS of
Eqn.~\ref{eqmath2} has most contribution from terms with smallest values
of  $(Z_i-Z')$ and
$(A_i-A')-(Z_i-Z')$. Note that we have chosen $Z_i>Z'$. For
most major QNSE elements we may then choose $A'=2Z'$ amd $A_i=2Z_i$.
For the SiG group pair \silicon[28] and \sulfur[32], Equation~(\ref{eqmath3}) becomes
\begin{eqnarray}
w = v Z_{\rm 32S}(1 - \ye - w)^2 (\ye - w)^2  ,
\label{eqmath4}
\end{eqnarray}
which is quartic in $w$. To order $\ye^2$, Equation~(\ref{eqmath4}) has the solution 
\begin{equation}
w= \frac{1}{2v Z_{\rm 32S}}\left [ \left (2\ye+1\right )\pm
\sqrt{\left(4\ye v Z_{\rm 32S}+1 \right )} \right ] 
\label{finalw}
\end{equation}
Substituting $Z_{\mathrm{32S}}=16$ and expanding the square root
leads to a zeroth order term that is a constant and a first order term
that is is linear in $\ye$.  From the expression for $w$ in
Eq. \ref{eqmath3}, the largest contribution comes from
$Z_{28Si}\ysi$. We thus identify $\ysi$ as the constant term and the
linear term in $\ye$ with $\ys$. Finally, we identify $\yca$ with
keeping the higher order terms in $\ye$ in Equation~(\ref{eqmath4}).

\section{Verification of QNSE from simulation models}
\label{sec:essential}
Here we 
 suggest
the trends predicted by our QNSE-based
theoretical relations are manifested in the abundances derived from
common 1D \SNIa\ models. In the next subsection we describe the
simulation model which we use to test our QNSE predictions.

\subsection{Description of simulation models}
We begin with the W7 model
\citep{nomoto_1984_aa,thielemann_1986_aa,iwamoto_1999_aa} because the
synthetic light cueves and spectra from W7
models have been extensively analyzed
\citep[e.g.][]{nugent_1997_aa,hachinger_2009_aa,jack_2011_aa,van-rossum_2012_ab}.
W7 is a 1D explosion model with a parameterized flame
speed that captures the stratified ejecta observed in Branch
normal \SNeIa\ \citep{branch_1993_aa}.  For our purposes, a model is
a series of temperature and density snapshots from ignition at $\textrm{time}=0\,\mathrm{s}$
to homologous expansion at $\textrm{time}=4.1\,\mathrm{s}$.  The W7 model assumes a solar
\neon[22] mass fraction uniformly distributed throughout the white
dwarf.  Our W7-like models change the assumed value of the
uniformly distributed \neon[22] mass fraction.  Specifically, we
control the value of the electron fraction $\ye$, which is set by the original composition and the pre-explosive
convective simmering \citep{timmes_2003_aa,piro_2008_aa,chamulak_2008_aa,townsley_2009_aa,walker_2012_aa}, 
by setting the mass
fraction of \neon[22]: $\ye=0.5 - X(\neon[22])/22=0.5 - Q \cdot
X(\neon[22])_{\odot}/22$, where $Q$ is a multiplier on the solar
\neon[22] mass fraction $X(\neon[22])$.

The nucleosynthesis of these W7-like models (see Figure
\ref{fig:freeze}) is calculated by integrating a 489 isotope nuclear
reaction network \citep{timmes_1999_ab} over the thermodynamic
trajectories of each Lagrangian mass shell.  This post-processing
calculation is not precisely self-consistent because the W7
thermodynamic trajectories have the built-in assumption of an energy
release from the original W7 model, carbon+oxygen material
complimented with a solar \neon[22] mass fraction.  While changes to
the abundance of \neon[22], hence $\ye$, slightly influence the energy
generation rate \citep{hix_1996_aa,hix_1999_aa}, burned material still
reaches QNSE conditions and our analysis should still hold.

\subsection{Verification of local QNSE relations}
\label{sec:abundance_trends}
First we test the validity of QNSE in the abundances 
synthesized in our W7-like \SNIa\ models. 
We use local values of $\rho$
and T from these models to construct the QNSE predicted abundances. 

\begin{figure}[htbp]
\centering{\includegraphics[width=6.5in]{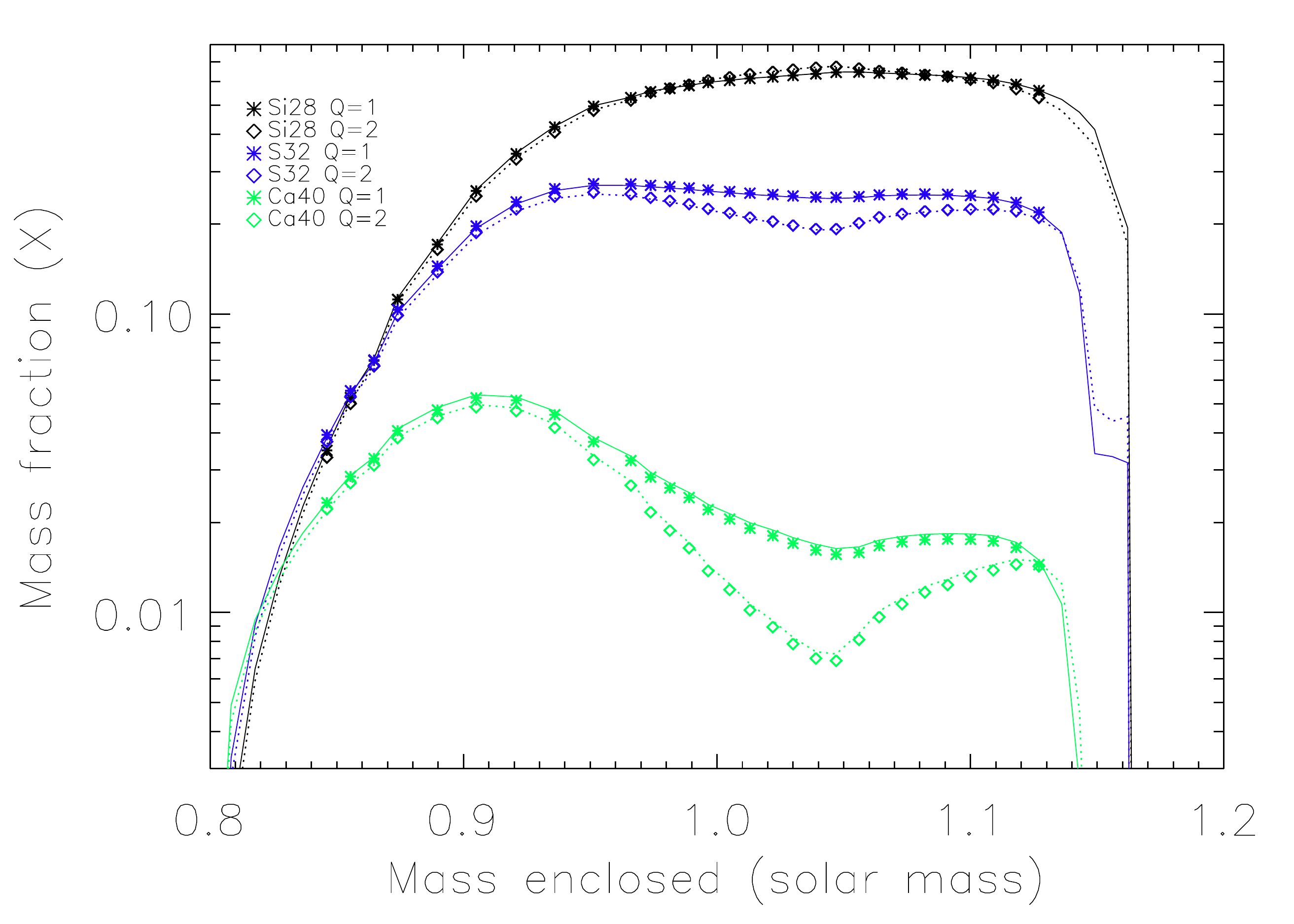}}
\caption{
Mass fractions of the major SiG elements as a function of interior
mass for the post-processed W7-like models (solid and dashed lines)
and the QNSE calculations (symbols) at $t=1.19\,\mathrm{s}$. Two cases are
shown, one for a solar \neon[22] abundance ($Q=1$) and one a twice
solar \neon[22] abundance ($Q=2$). Overall, the agreement between the
post-processed W7-like models and the analytical QNSE results are
satisfactory. \calcium[40] shows the largest change, up to a factor of two,
while \silicon[28] is insensitive to changes in \neon[22].
}
\label{f:ysi}
\end{figure}

\begin{figure}[htbp]
\centering{\includegraphics[width=6.5in]{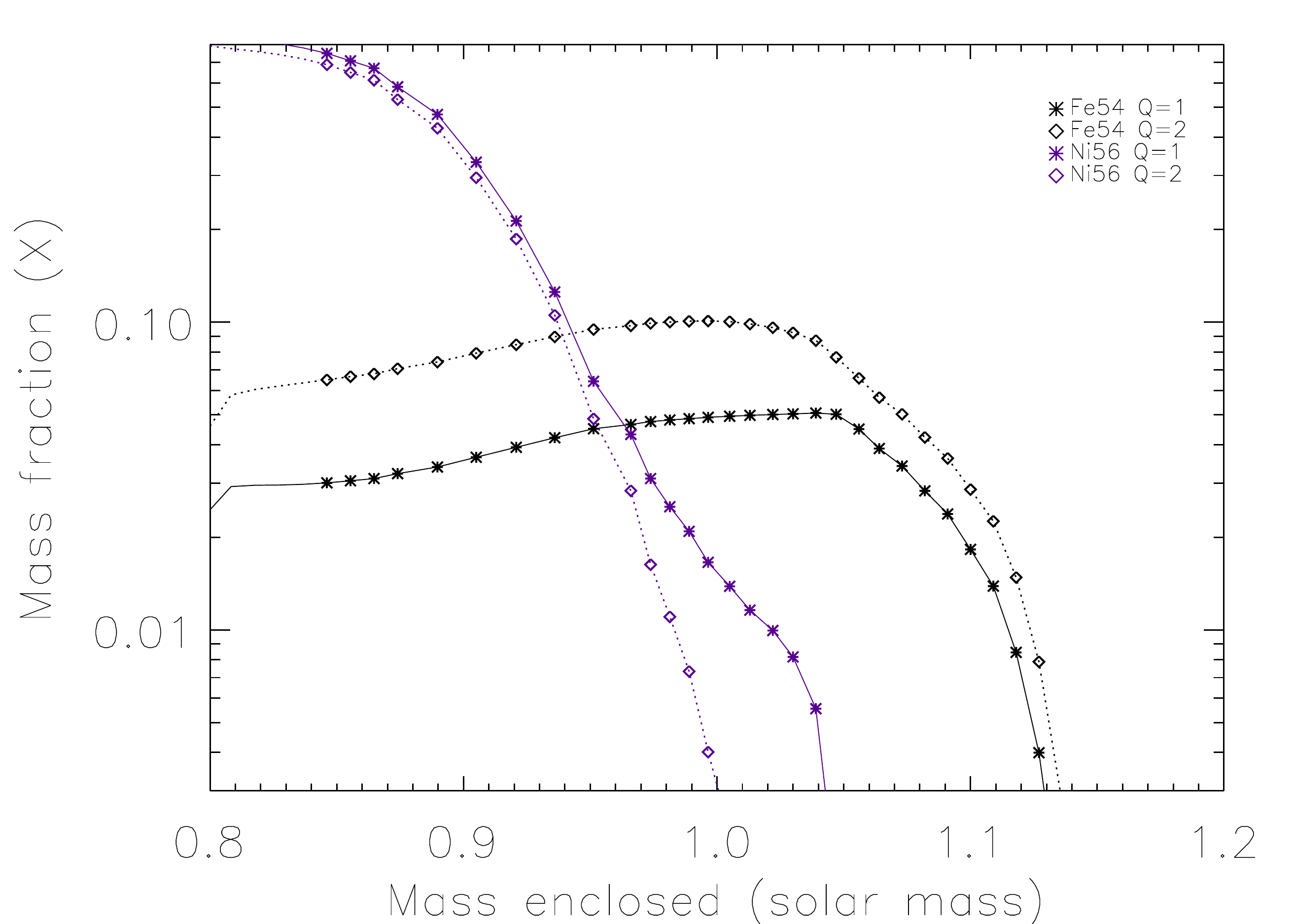}}
\caption{
Same as Figure \ref{f:ysi} but for the major FeG elements.  \iron[54]
shows the largest change with changes in \neon[22].  Note that
\iron[54] is the only significant iron isotope present when the SiG
elements of Figure \ref{f:ysi} are dominant.
}
\label{f:yfe}
\end{figure}

Figure \ref{f:ysi} shows the final mass fractions of the SiG elements
\silicon[28], \sulfur[32], and \calcium[40] between 0.8\,\msun\ and
1.2\,\msun. Two cases for each element are shown, $Q=1.0$ and 2.0,
representing 1.0 and 2.0 times the solar \neon[22] abundance,
respectively. Solid lines represent the results from post-processed
W7-like models and the symbols represent the results from our 
QNSE solutions.  Figure \ref{f:yfe} shows the final mass fractions of
FeG elements \iron[54] and \nickel[56] over the same mass range and
white dwarf metallicities as in Figure \ref{f:ysi}.  These figures show
\calcium[40] and \iron[54] have the largest systematic changes (up to
a factor of two depending on the mass shell) within the silicon-rich
region as the electron fraction varies. Also, a major conclusion from
Figure \ref{f:yfe} is that \iron[54] is the $only$ iron isotope and the
most abundant FeG element in the QNSE regime bound by regimes enclosing
0.8$\msun$ and 1.1$\msun$. This relation is also supported by
\cite{mazzali_2013_aa}.

A weakly varying $\Phi$ (see Eq.~(\ref{e:phi})) implies $d\Phi/dT \simeq 0$.  It follows that the
constituent ratios ${\ysi}/{\ys}$ and ${\ys}/{\yca}$ reach an extremum
at the same temperature. This property is evident in Figure \ref{f:ysi}:
when there is a deflection in ${\ysi}/{\ys}$ with mass coordinate
there is a corresponding variation in ${\ys}/{\yca}$.  In addition, the individual abundances $\ysi$, $\ys$, and
$\yca$ reach an extremum at the same ($\rho$,$T$) point where the
ratios reach an extremum.  In general, isotopes with $A=2Z$ will have
a QNSE abundance $\ya$ that scales as $(\yp\yn)^{n}$, where $n$ is a
positive or negative integer. These isotopes can be expressed as a
function of ${\ysi}/{\ys}$ (Eq.~(\ref{e:phi}) is one example) and
will reach an extremum at the same ($\rho$,$T$) point as $\ysi$.  The
only assumption in deriving these properties is that the system
achieves QNSE conditions, where most of the SiG elements are
synthesized.  The values of the ratios ${\ysi}/{\ys}$ and
${\ys}/{\yca}$ at the extremum will, of course, depend on the details
of the explosion.  Treating $\Psi$ (see Eq.~(\ref{e:psi})) as a
quasi-constant and following arguments similar for the SiG elements,
one finds $\yfe$, $\ynia$ and $\ysi$ also reach an extremum at the
same ($\rho$,$T$) point, as shown in Figure~\ref{f:yfe}.
\subsection{Global abundances as predicted by QNSE}

Next we explore the global  abundances of the SiG and
FeG elements with $\ye$ when our W7-like models reach homologous
expansion at $t=4.1\,\mathrm{s}$.  Figure~\ref{f:nucleo} shows the total \silicon[28],
\sulfur[32], and \calcium[40] molar abundances ejected as a function
of the $\ye$.  The curves correspond to post-processing the W7-like
thermodynamic trajectories and the symbols are the results from our
analytical QNSE model.  Both the post-processing and the model
independent QNSE results suggest a nearly constant \silicon[28] yield
with respect to $\ye$, a systematic quasi-linear \sulfur[32] yield
with respect to $\ye$, and a more complex trend for the global
abundance of \calcium[40] with $\ye$. Among the SiG elements,
\calcium[40] has the largest sensitivity to the electron fraction, in
agreement with the trends seen in the local abundances.  These results
are in accord with trends explored in \S\ref{sec:trends}.

\section{Possible application to observations}

That QNSE abundance ratios are manifest in our W7-like post-processing
models suggests the QNSE relationships may be applicable to
observations. We seek connections between abundances derived from a
SNIa spectra that can be mapped to a $\ye$ of the silicon-group
ejecta. The first step to such a mapping involves the most important
test, whether or not the derived abundances of major elements are in
QNSE.  In other words, are the abundances levels produced in QNSE retained
duing the subsequent freeze-out?

\subsection{Verification of QNSE at freeze-out from simulation models}
\begin{figure}[htb]
\centering{\includegraphics[width=6.5in]{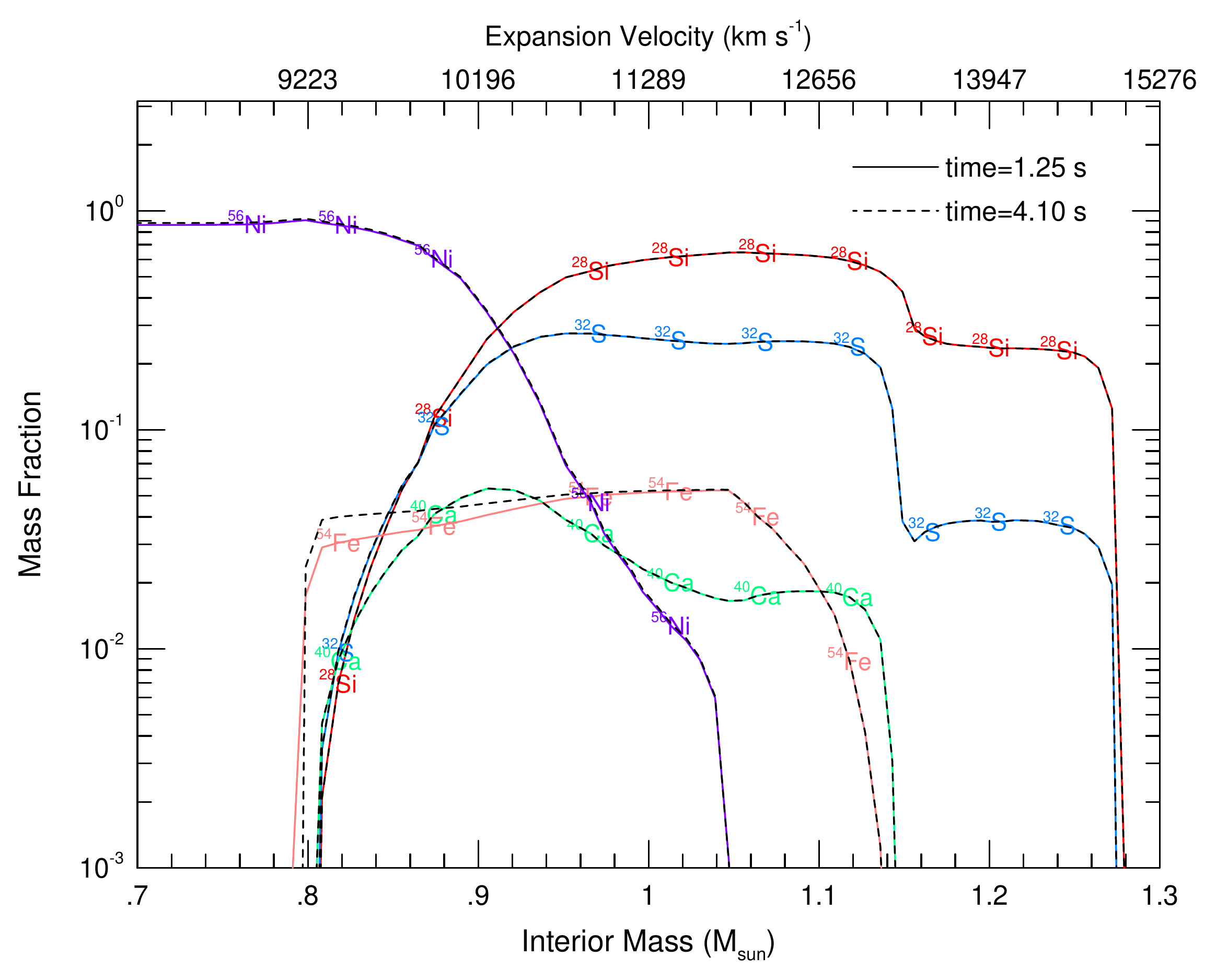}}
\caption{
Mass fractions of the major elements during a W7-like explosion. Solid
colored lines correspond to $t=1.25\,\mathrm{s}$ and and dashed black lines
correspond to $t=4.10\,\mathrm{s}$ since ignition. The abundances produced by
QNSE conditions at $t=1.25\,\mathrm{s}$ are the same as the final abundances
when nuclear reactions freezeout at $t=4.0\,\mathrm{s}$. The expansion
velocities on the upper $x$-axis are when the W7-like model
reaches peak luminosity.
}
\label{fig:freeze}
\end{figure}

\begin{figure}[htb]
\centering{\includegraphics[width=4.0in]{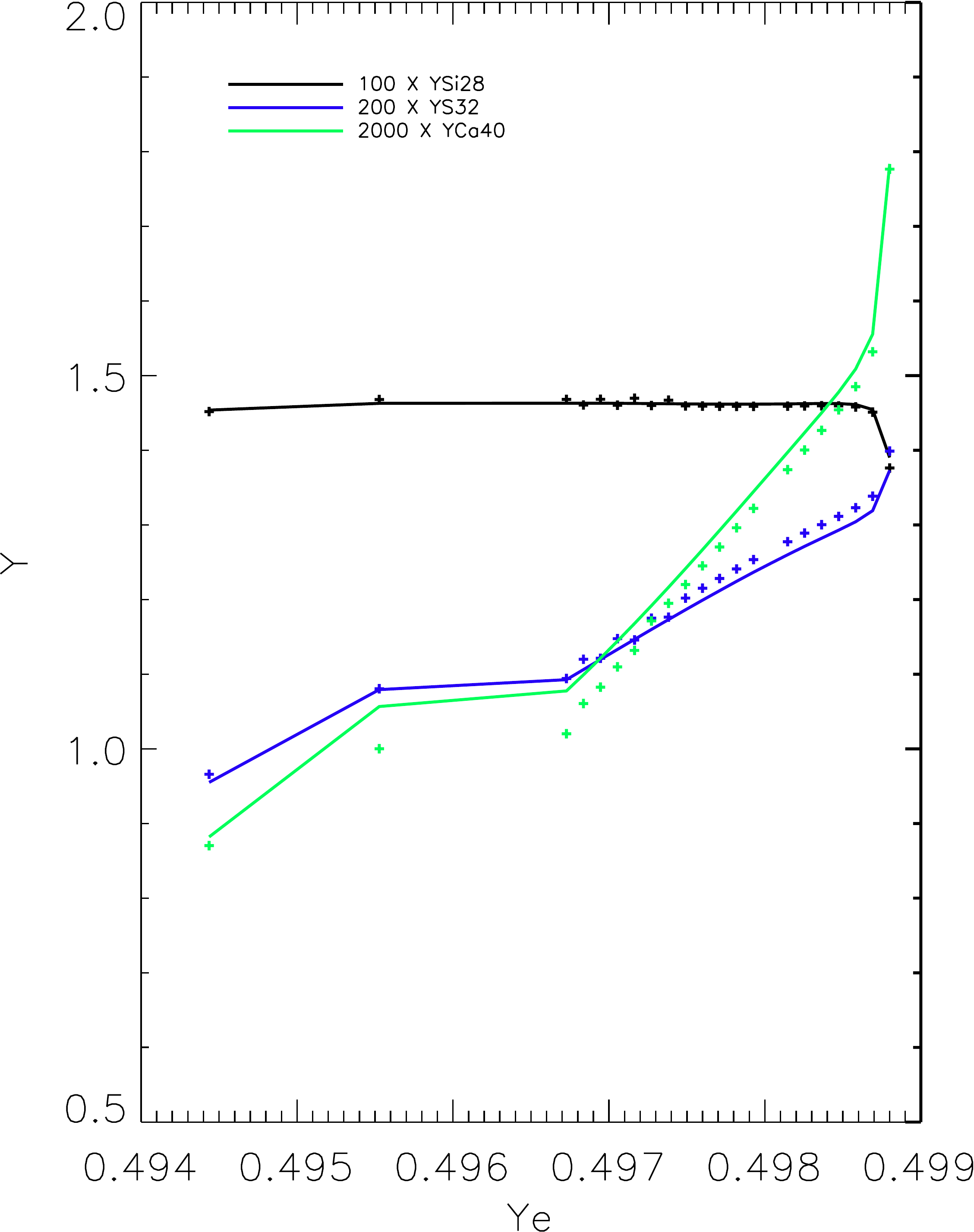}}
\caption{
Global abundances of \silicon[28], \sulfur[32], and \calcium[40] as a
function of the electron fraction $\ye$ produced by the post-processed
W7-like models (lines) and the analytical QNSE results (symbols).  As
for the trends in the local abundances (Figs.~\ref{fig:freeze}--\ref{f:yfe}), \silicon[28] is independent of $\ye$, \sulfur[32] shows
a near linear dependence, and \calcium[40] shows a more complex, but
near quadratic, dependence with $\ye$.
}
\label{f:nucleo}
\end{figure}

Figure \ref{fig:freeze} shows the local abundances
of the major SiG and FeG elements between mass shells $0.7\,\msun$ and
$1.3\,\msun$ at $t=1.125\,\mathrm{s}$ (solid colored lines) and $t=4.10\,\mathrm{s}$ (dashed
black lines) in one of our W7-like models.  At $t=1.25\,\mathrm{s}$ the burning front has just passed over
the $1.28\, \msun$ mass coordinate and most of the synthesized \silicon[28],
\sulfur[32], and \calcium[40]  have reached their
equilibrium abundances.  In this mass region, the peak temperatures,
($3\textrm{--}5)\times10^9\,\mathrm{K}$, and peak densities, $(2\textrm{--}4)\times10^7\,\mathrm{g\,
cm^{-3}}$, ensure QNSE conditions
\citep{nomoto_1984_aa,thielemann_1986_aa}.  The choice of $t=1.125\,\mathrm{s}$
is arbitrary and can be replaced by any epoch in any model when the
material reaches QNSE conditions.  At $t=4.10\,\mathrm{s}$ the explosion has
entered homologous expansion and synthesis of all the elements has
stopped due to the decreasing temperature.  Complete freezeout does
not occur for \iron[54] interior to $0.95\msun$ at $t=1.25\,\mathrm{s}$ due to
residual weak reactions.  Figure \ref{fig:freeze}
 suggests
that abundances generated when QNSE conditions apply are preserved during
the subsequent freezeout.  The abundance levels at this epoch may be
reflected in the observed spectra over subsequent days.  Therefore,
applying the QNSE equations to recover $\ye$ from the major SiG and FeG elements
as we have done in \S\ref{sec:constructing} is justified.

\subsection{Expected change in spectra due to change in $\ye$}
\label{sec:spectra}
The next step to mapping the observed abundances into $\ye$ at
explosion involves estimating the change in flux and luminosity as
$\ye$ changes.  Its important to estimate this change which dictates
the level of resolution in $\ye$ mapping accessible from the observed
abundances, if determined accurately from the spectra. There are
several complicating factors that contribute to estimating acurate
abundances from an observed spectra.  

Below we estimate the change
in flux with respect to $\ye$ using synthetic spectra from 
radiative transfer modeling.
 
We use the PHOENIX radiation transfer code
\citep{hauschildt_1997_aa,jack_2009_aa,jack_2009_ab,jack_2011_aa} to
produce synthetic spectra from our W7-like models.
The thermodynamic profiles of the W7-like models end when the
explosion reaches homologous expansion, about $4\,\mathrm{s}$ after ignition.  The
density, velocity, and abundance profiles are then homologously and
adiabatically expanded to 5 days after the explosion using analytical
expressions that account for the local decay of \nickel[56] and
\cobalt[56]. This is a reasonable assumption for \SNIa \ after the
initial break out \citep{arnett_1982_aa}.  From day 5 onwards, we
address LTE radiative transfer through the expanding remnant in 0.5
day increments to about 21 days to calculate synthetic spectra.
At each of these 0.5 day increments, we solve an energy
equation that includes the contribution of the adiabatic expansion,
the energy deposition by $\gamma$-rays, and absorption and emission of
radiation. As a result we always obtain a model atmosphere that is in
radiative equilibrium. 

Figure \ref{f:spectra} shows the synthetic spectra at day 15, near
peak luminosity, for the W7-like models with $Q=0.0$, 1.0, 2.0, and
4.0, representing 0.0, 1.0, 2.0, and 4.0 times the solar \neon[22]
abundance, respectively.  Changes in the synthetic spectra shown in
Figure \ref{f:spectra} are due to changes in the abundance of a given
element.  The SiG material has post-explosion homologous expansion
velocities ranging between $9000\textrm{--}13000\,\mathrm{km\,s^{-1}}$
\citep[e.g.][]{nugent_1997_aa,hachinger_2009_aa,jack_2011_aa,van-rossum_2012_ab}.
The continuum
optical depth of the SiG material is between 0.2--0.8, which was
evaluated from the continuum opacity at $5500\,\angstrom$ at peak luminosity.
The wavelength resolution of the PHOENIX calculation is set to $1\,\angstrom$ and 
the distance for the flux scale shown on the $y$-axis is the
velocity of the outermost expanding shell times the time since the
explosion. The PHOENIX model uses complete redistribution, with the
same emission and absorption profile function with respect to the
frequency spread around the center of the line frequency.  Figure
\ref{f:spectra} suggests that the Ca II feature 
changes the most with $\ye$ when compared to S II and Si II
features.  The
Si II line changes the least with variations in $\ye$. These trends,
Ca being the most sensitive, S having a near linear dependence, and Si
the least sensitive, are a reflection of the nucleosynthesis trends
shown in Figure \ref{f:nucleo}. We stress the LTE synthetic spectrum 
of our W7-like models is not compared to observational data, and thus 
our W7-like models may not accurately model real \SNIa.

\section{Discussion}
\label{sec:implications}

\begin{figure}[htb]
\centering{\includegraphics[width=6.5in]{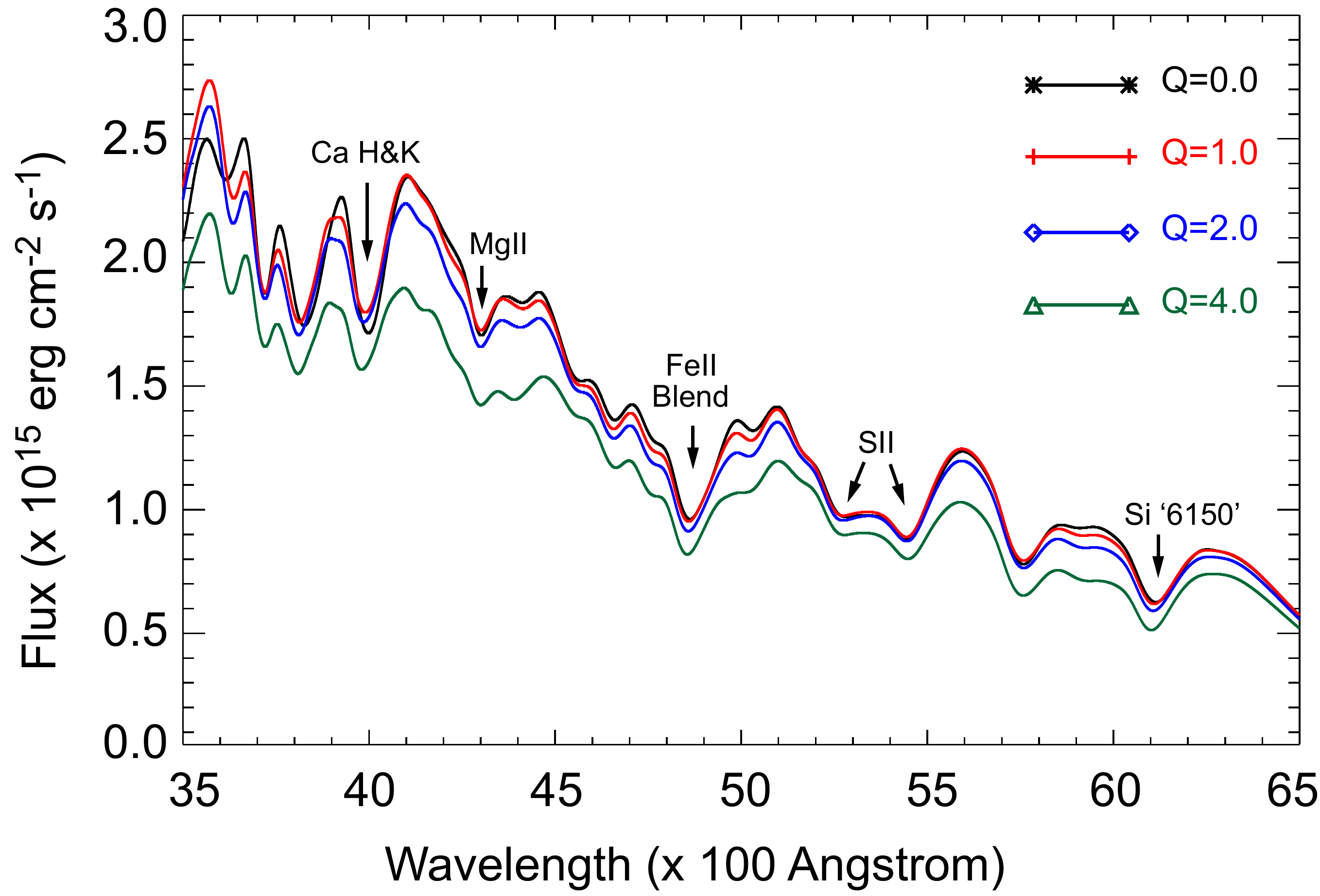}}
\caption{Synthetic spectra for the W7-like models with 0.0. 1.0, 2.0, and 4.0 times
the solar \neon[22] abundance. Key abundance features are labeled.}
\label{f:spectra}
\end{figure}

We summarize our findings and conclusions of this work below.
\begin{enumerate}

\item 
We construct a mapping between $\ye$ at explosion and abundances of a
few major elements from SiG and FeG. Specifically, we show that the
abundance $\ysi$ and the abundance ratios $\ys/\ysi$, $\yca/\ys$, and
$\yfe/\ysi$ describe the complete basis to reconstruct the $\ye$
 of silicon group material at explosion. This conclusion
simply follows from the QNSE relations.   If these four 
abundances are determined exactly, they are sufficient
to recover $\ye$ to 6\%. This is because these isotopes dominate the
composition of silicon-rich material and iron-rich material in QNSE.
\item From the widely successful Our W7-like simulations 
 suggest that the
major element abundances manifest QNSE trend at freeze-out. This
 might imply the abundances underlying an observed spectra 
obey  the QNSE relations. This is one of our most important 
chief findings, which  may allow
one to a mapping of the major elements found determined
from an observed spectra onto $\ye$  of the SiG material at explosion.

\item We find that among the major SiG
elements, Ca abundance is expected to change the most with respect to
$\ye$ at explosion. This fact follows from  QNSE relations and
manifest in the W7-like simulations as well.  The Si
abundance is not expected to change much as $\ye$ is varied. We
predict that in the QNSE regime $\yfe$ is the only isotope of iron and
the only element in  the FeG to have an abundance comparable
with respect to  the
SiG elements. This conclusion also follow from QNSE equations.

\item Fig.~\ref{f:spectra} gives the reader a rough idea
estimate of the change in
spectral features with respect to $\ye$. These spectral
features are driven by abundances that follow from QNSE.  We
conclude from the calculated synthetic spectra, A flux resolution of
0.1$\times$ 10$^{15}$ erg cm$^{-2}$ s$^{-1}$, if achieved , corresponds
to a $\delta \ye=0.002$ resolution.  This estimate is under the
assumption that abundances can be determined precisely 
 very accurately from an
observed spectra  and radiative transfer modeling. 
One of the avenues of recovering the abundance
from a spectra is radiative transfer modeling of the
spectra. Radiative transfer modeling treats the line formation in
detail but is challenging and involves some modeling parameters.
\end{enumerate}

\clearpage
\acknowledgements 
We thank Friedel Thielemann for providing the initial W7 thermodynamic
trajectories. We also thank the  anonymous reviewer for 
several suggestions which  significantly
improved the manuscript. This work was supported by the NSF through grant AST
08-06720 (FXT) and PHY 08-022648 for the Frontier Center ``Joint
Institute for Nuclear Astrophysics'' (JINA), by NASA through grant
NNX09AD19G, and by the DOE through through grants DE-FG02-07ER41516,
DE-FG02-08ER41570, DE-FG02-08ER41565 and DE-FG02-87ER40317.  SD
received support from a SESE Postdoctoral fellowship at Arizona State
University.

\clearpage

% Figures go here 

\clearpage

\clearpage

\clearpage

\clearpage

\clearpage

% Bibliography goes here

% \bibliographystyle{apj}

%\bibliography{fxt_master}

\end{document}